 \documentstyle[12pt]{article}

\begin{document}
%%%%%%%%%%%%%%%%%%%%%%%%%%%%%%%%%%%%%%%%%%

\def \adss {$AdS_5 \times S^5$\ }
\def\aA{{A}}
\def\aB{{B}}
\def\aC{{C}}
\def\aE{{E}}
\def \om {\omega}

\def \N {{\cal N}}
\def \lc {light-cone\ }
\def \ta { \tau}
\def \s { \sigma }
\def  \gg  { {\rm g}}
\def \sg {\sqrt {g }}
\def \te {\theta}
\def \vp {\varphi}
\def \gij {g_{ab}}
\def \xp {x^+}
\def \xm {x^-}
\def \p {\phi}
\def \vt {\theta}
\def \bx {\bar x} \def \a { \alpha}
\def \r {\rho}
\def \fourth {{1 \ov 4}}
\def \DD {{\cal D}}
\def \half {{1 \ov 2}}
\def \inv {^{-1}}
\def \ri {{i}}
\def \D {{\cal D}}
\def \DD {{\rm D}}
\def \vr {\varrho}
 \def \diag {{\rm diag}} \def \td { \tilde }
\def \tta {\td \eta}
\def \cA {{\cal A}}
\def \cB   {{\cal B}}
\def \na {\nabla}
\def \te{\theta}
\def \t {\theta}
\def \la {\label}
\def \alpr {\alpha'}
\def \bepr{\beta'}
\def \del{\partial}
\def \m {\mu }
\def \n {\nu }
\def \ha { { 1 \over 2}}
\def \cN {{\cal N}} 
\newcommand{\eq}[1]{Eq.~(\ref{#1})}
\newcommand{\rf}[1]{(\ref{#1})}

\def \four{{\textstyle {1\ov 4}}}
 \def \third { \textstyle {1\ov 3}}
\def\det{\hbox{det}}
\def\be{\begin{equation}}
\def\ee{\end{equation}}
\def \ci {\cite}
\def \g {\gamma}
\def \G {\Gamma}
\def \k {\kappa}
\def \l {\lambda}
\def \L {{\cal L}}
\def \Tr {{\rm Tr}}
\def\apr{{A'}}
\def\bpr{{B'}}
\def \CC {{\cal C}}
\def\vm{{\mu}}
\def\vn{{\nu}}
\def \td {\tilde }
\def \b{\beta}
\def \gg {{\cal g}}
\def \R {R^{(2)} } 
\def \foot {\footnote}
\def \bi{\bibitem}
\def \la {\label}
\def \tr {{\rm tr}}
\def \ha {{1 \over 2}}
\def \ep{\epsilon}
\def \CC{{\cal C}}
\def \ov {\over}
\def \JJ {{\cal J}}
\def \z {\zeta}

\hoffset=-25pt
\voffset=-1.5cm
\textwidth=15.8cm
\textheight=23cm

\catcode`\@=11
%%%%%%%%%%%%%%%%%%%%%%%%%%%%%%%%%%%%%%%%%%%%%%%%%%%%%%%%%%%%%%%%%%%%%%%%%%
%%%   paper [12 Pages]
%%%%%%%%%%%%%%%%%%%%%%%%%%%%%%%%%%%%%%%%%%%%

\begin{titlepage}
\begin{flushright}
hep-th/0008107\\
\end{flushright}
\vspace{.5cm}

\begin{center}
{\LARGE ``Long" Quantum Superstrings in $AdS_5 \times S^5$ } \\[.2cm]
\vspace{1.1cm}
{\large  A.A. Tseytlin${}^{{\rm a,}}$\footnote{\ Also at Blackett
Laboratory,
Imperial College, London and   Lebedev Physics Institute, Moscow.\
\ \
 \ \ \ \ \ \ E-mail: tseytlin@mps.ohio-state.edu} }\\

\vspace{18pt}
 ${}^{{\rm a\ }}${\it
 Department of Physics,
The Ohio State University  \\
Columbus, OH 43210-1106, USA\\
}
\end{center}

\vspace{2cm}

\begin{abstract}
%{\bf Abstract} \end{center}

We discuss the  computation of  quantum corrections near  long 
 IIB superstring configurations in \adss space 
which are 
related  to the Wilson loop expectation values in 
the strong coupling expansion of
the dual  $\N=4$ SYM theory  with  large $N$.
We use  the Green-Schwarz  description of 
superstrings in  curved R-R  backgrounds
and demonstrate that it 
is well-defined and useful 
  for developing perturbation theory 
near long string background. 

\end{abstract}

\end{titlepage}
\setcounter{page}{1}
\renewcommand{\thefootnote}{\arabic{footnote}}
\setcounter{footnote}{0}

%%%%%%%%%%%%%%%%%%%%%%%%%%%%%%%%%%%

\section{Introduction}
%%%%%%%%%%%%%%%%%%%%%%%%%%%%%%%%%%%%%

Further  progress in  clarifying   the relation 
between strings and large $N$ (non) supersymmetric 
gauge theories 
depends on better understanding of superstrings 
in $AdS$-type  spaces with Ramond-Ramond (RR)  fluxes
\ci{POL,mald}. 
The simplest most symmetric example is provided by \adss
background of type IIB superstring theory.
 One could hope that this theory,
though quite nontrivial being dual to large N 
\ $\N=4$  Super Yang-Mills theory \ci{mald}, 
should be, like the flat superstring,   explicitly solvable. 

To be able to  treat this theory in an efficient 
way one is to use the Green-Schwarz (GS)  approach \ci{GS}
where one views the string as moving in a superspace 
$(x^m,\theta^I_\a)$ with $\theta$'s being space-time spinors.
Then  the coupling of the string  to 
RR  background has a local form  $ \bar \te \Gamma^{...} \te
\del x \del x  F_{...}$, just like its coupling 
to the curvature. It is useful to  note  
that while the  string ``feels"  the  metric already at the classical
level, 
it interacts with  the RR background  only through its quantum
fluctuations. 

 There seems to be no alternative to the GS-type 
description: even the simplest one-loop computation 
in GS formalism like the one described  in \ci{DGT}
would require a summation of infinite number of diagrams with 
any number of flat-space  RR vertex  operator insertions
if done in NSR formalism.
In fact, the use of GS action is very natural  from 
 ``solitonic" point of view. 
If one  views fundamental strings as ``electric"-type 
objects appearing  in $D=10$  supergravity, the 
effective action for collective coordinates 
of a long such string  will  naturally have 
a GS form,   as dictated by the residual space-time 
(super)symmetries
(the characteristic  non-linear  $\del x \bar \te \Gamma^{...} \te$
kinetic term 
originates from the standard Dirac $D=10$ fermion action 
reduced to the world surface  of the string, with Dirac 
matrices becoming the  projected ones).
The theory should of course contain not only long but also short
strings which should also be described by the same manifestly
space-time supersymmetric action.

A simple way to  construct the flat space GS action is 
to view  strings as propagating in flat coset 
superspace =
(10-d super Poincare group)/( 10-d  Lorentz group).
The action is then a kind of Wess-Zumino-Witten 
action which has global supersymmetry and local
$\k$-symmetry  which ensures the correct number of the fermionic
degrees of freedom.
 It is defined in terms of the 
 basic objects 
which are the left-invariant Cartan  1-forms 
on the
type IIB coset superspace
$
G^{-1}dG =
L^{\hat{A}}P_{\hat{A}} + L^I Q_I\ , $  $
 L^{\hat{A}} = dX^M L^{\hat{A}}_{M}, $ $ 
X^M=(x, \theta)  , $ 
where  $ [P_{\hat A}, P_{\hat B}]=0\,, $  $ 
\  \{ Q_I, Q_J\}
=-2{ i}  \delta_{IJ} (\CC\Gamma^{\hat A}) P_{\hat A}\,$
and  
$G= G({x,\theta})$ is a coset representative
$
G({x,\t}) = {\rm exp} ( x^{\hat A} P_{\hat A}  + \t^I Q_I)$.
The coset space vielbeins 
are given
by
$
L^{\hat A} = d x^{\hat A}  - i \bar \t^I \G^{\hat A}d\t^I \,, $ $ 
L^I = d \t^I. $ 
Here  $  x^{\hat A} $ are flat bosonic coordinates and 
 $\theta^I$ ($I=1,2$)  are two left Majorana-Weyl 10-d spinors.
The general form of the action is \ci{HM}
\be
S=-\frac{1}{2}\int_{\del {M_3}}  d^2\sigma\  \sqrt{g}\
g^{\vm\vn}\
 L_{\vm}^{\hat A}  L_{\vn}^{\hat A}
+  { i}\int_{M_3} {\cal H} \ , \ \ \ \ \ \ 
{\cal H} = 
 s^{IJ}   L^{\hat A}\wedge  (\bar{L}^I\Gamma^{\hat A}
\wedge  L^J)
\ ,
\la{actif}
\ee
where   $s^{IJ}\equiv {\rm diag}( 1,-1)$, 
 ${ 2\pi \alpha'}=1$ and $d {\cal H}=0$.  The 2-d metric
$g_{\vm\vn}$ ($\vm,\vn=0,1$)  has  signature
$(-,+)$, and $g\equiv - \det g_{\vm\vn}$.
The first ``kinetic" term in the action 
has ``degenerate" metric, containing only the square of
translational
but not spinor  Cartan forms, as the latter would 
lead to an  action  quadratic in fermionic derivatives and thus to 
potential non-unitarity. 
The explicit 2-d  component form of the GS action is  \ci{GS}
$$
S_0=
  \int d^2 \sigma\  \bigg[ - \ha
\sqrt{g} g^{\vm\vn} (\del_\vm x^{\hat A} -
{ i} \bar \t^I \G^{\hat A} \del_\vm \t^I)
(\del_\vn x^{\hat A} -
{ i} \bar \t^J \G^{\hat A} \del_\vn \t^J)
$$ \be
-{i} \ep^{\vm\vn} s^{IJ}   \bar \t^I \G^{\hat A} \del_\vn 
\t^J
(\del_\mu  x^{\hat A} -  \ha {i}   \bar \t^K \G^{\hat A} 
\del_\vm \t^K)
 \bigg] \ .
\la{GRE}
\ee
The  key observation that allows one to construct the GS action 
for superstrings in \adss is that this  bosonic 
space is   a coset

 $G/H=SO(2,4)/SO(1,4) \times SO(6)/SO(5)$
 
\noindent 
and  also has maximal supersymmetry. 
As a result,  one can consider strings moving not 
in flat superspace  but in supercoset space with 
the right bosonic  part and right number
supersymmetries: one is to replace 
the 10-d super Poincare group  by  a smaller one $PSU(2,2|4)$
and  the Lorentz group $SO(1,9)$  by its subgroup 
 $SO(1,4) \times SO(5)$, i.e. to consider the supercoset 
$PSU(2,2|4)/[SO(1,4) \times SO(5)]$ \ci{MT},
which still has 10 bosonic and 32   Grassmann 
dimensions.

The bosonic part of the GS action is 
simply the standard symmetric space sigma model
$L= \Tr [ (g^{-1} \del g)_{G/H} ]^2$.
This is obviously not a conformal sigma model;
it is the 
addition of the fermions $\te$ that  converts it into a 
 conformal 2-d  model of a novel type. The fermions couple
the $AdS_5$ and $S^5$ parts of the action together  and
contribute the RR 5-form  term to the beta-function
of the metric  making it  vanish \ci{MT}, 
$R_{mn} - (F_5 F_5)_{mn} =0$.

The  superalgebra $psu(2,2|4)$ 
plays the
central  role in the construction of the GS action in \adss
\ci{MT}.
Its even part 
   is the sum of  the algebra $so(4,2)$ which is the
isometry algebra of $AdS_5$ and the algebra $so(6)$ which is the 
isometry
algebra of $S^5$.
 The odd part  consists of 32 supercharges
corresponding to  32 Killing spinors in $AdS_5\times S^5$
vacuum \ci{john}
of type IIB supergravity
 \cite{GM1}. 
In the ``5+5'' split  its generators  are:
$( P_A,J_{AB}; \  P_{A'},J_{A'B'}; \ Q_{\a\a' I})$, i.e. the 
the  two sets of translations and rotations in $so(2,4)$ 
and $so(6)$  ($A=0,1,...,4;\  A'= 1,...5$)
 and  32  supercharges ($ \a=1,2,3,4; \ \a'=1,2,3,4$).
 The commutation
relations of $psu(2,2|4)$ in $so(4,1)\oplus so(5)$
basis   are \ci{MT}
$
[{P}_\aA,{P}_\aB]={J}_{\aA\aB}\,, $ $ 
\ 
[P_\apr, P_\bpr]=-J_{\apr\bpr}\,, $ $ 
\ \ 
[J^{\aA\aB},J^{\aC\aE}]=\eta^{\aB\aC}J^{\aA\aE}+..., $ $ 
\ 
[J^{\apr\bpr},J^{C'E'}]=\eta^{\bpr C'}J^{\apr  E'}+...,  $ $  
[Q_{I},{P}_\aA]
=-\frac{{i}}{2}\epsilon_{_{IJ}}Q_{J}\gamma_\aA \,, $ $ 
\ 
[Q_{I},{J}_{\aA\aB}]=-\frac{1}{2} Q_{I}\gamma_{\aA\aB}\,, $ $ 
\  
[Q_{I},P_\apr]
=\frac{1}{2}\epsilon_{_{IJ}} Q_{J}\gamma_\apr\,, $ $ 
\ \ 
[Q_{I},J_{\apr\bpr}]=-\frac{1}{2}Q_{I}\gamma_{\apr\bpr} , $
and $$
\{Q_{\alpha \alpr I}, Q_{\beta \bepr J}\}
=\delta_{_{IJ}}
[-2{i}C_{\alpr\bepr}(C\gamma^\aA)_{\alpha\beta} 
{P}_\aA 
+ 2C_{\alpha\beta}(C^\prime\gamma^\apr)_{\alpr\bepr}P_\apr]  $$
 \be + \ \epsilon_{_{IJ}}
[C_{\alpr\bepr}(C\gamma^{\aA\aB})_{\alpha\beta} {J}_{\aA\aB}
-C_{\alpha\beta}(C^\prime\gamma^{\apr\bpr})_{\alpr\bepr}
J_{\apr\bpr}]
\ , \ee
where $C,C'$ are charge conjugation matrices.
Then 
$ G^{-1}dG
={L}^\aA {P}^\aA+\frac{1}{2}\hat{L}^{AB}\hat{J}^{\aA\aB} $ $ 
+L^\apr P^\apr+\frac{1}{2} L^{\apr\bpr} J^{\apr\bpr}
+ L^{I\alpha \alpr}Q_{I\alpha\alpr}\ ,$
and the coset representative can  be parametrised as 
$G(x,\te) = g(x) \ {\rm exp} ( \t \cdot Q) $.

 Following the same steps as  above in 
 the  construction 
of the flat space GS action  one finds \ci{MT}
that there exists 
a  unique  action of the type 
\rf{actif}  which has the required properties:	
its bosonic part is sigma model on \adss, 
it has  local $\k$-symmetry and global $PSU(2,2|4)$ 
symmetry,  and it reduces to the standard GS action in the flat
space limit.
The action has the same structure as \rf{actif} but now in terms of the 1-forms
corresponding to 
the  supercoset $PSU(2,2|4)/[SO(1,4) \times SO(5)]$. 
The leading terms in the resulting action are essentially the 
covariantisation
of the flat space action 
with   the  ordinary  derivatives on $\te$'s  replaced by 
generalized  \adss covariant ones
(containing extra terms linear in Dirac matrices that correspond to the 
interaction with RR background).

As in the standard  WZW model case 
it is  possible to argue that this action defines 
a conformal sigma model \ci{MT,MTOL}: (i) the WZ term in the action is not renormalized 
since the divergences are local  and manifestly covariant, while
the WZ term is not; (ii)  the kinetic $L^2$ term 
in the action can be renormalized only by an overall constant
because of the global symmetries of the sigma model;
(iii) assuming that $\kappa$-symmetry is preserved by regularization, 
it should relate the coefficients of the kinetic andthe  WZ terms
as it does at the classical level (i.e. $\kappa$-symmetry
plays  the role of the affine symmetry in WZW model).

The explicit form of
all  higher order terms in $\te$ in the action of  \ci{MT} 
can be found in 
\ci{KRR,MMT}. Fixing a ``covariant" 
$\k$-symmetry gauge (e.g. $\te^1 = \G_{0123} \te^2$) 
one  gets a relatively
simple  action  which  contains terms of quadratic and quartic
order in fermions only  \ci{pessan,kalram}.
The resulting action has the same feature
 as the flat space GS action: its fermionic 
 kinetic term   is coupled to derivative of  all bosonic string coordinates.
 It is non-degenerate when expanded  near ``long" string
 configuration \ci{KT}, and thus  the GS action 
 provides a well-defined and useful tool for computing quantum
 corrections to ``long" string configurations 
 ending on Wilson loops \ci{malda} 
 at the boundary of $AdS_5$ 
 \ci{KT,Forste,DGT}.
 
 The bosonic part of the action for a string in \adss\ 
is the sigma model corresponding to 
 the space-time metric
($m=1,\dots,4$)
\be 
ds^2=R^2 \bigg[ {1\over
w^2}\left(dw^2+dx^mdx^m\right)+d\Omega_5^2\bigg]
\,.\ee 
The  $\a'$ expansion
corresponds to the  large 't Hooft coupling $\l$ 
expansion in the dual gauge theory, with the leading classical 
contribution  proportional to ${R^2\ov \alpha'}
= \sqrt\lambda$ \ci{malda}.

%%%%%%%%%%%%%%%%%%%%
\section{Quadratic  approximation}
%%%%%%%%%%%%%%%%%%%%%%%%%%%%%%%%%%%%

The part of the action in
\ci{MT} quadratic in $\vt^I$ is a direct generalization of the
quadratic
term in the flat-space GS action  \rf{GRE}
\be 
S^{(2)}_{F} = {{i }} \int d^2 \s ( \sqrt{g}
g^{\vm\vn}\delta^{IJ} - \ep^{\vm\vn} s^{IJ} ) \bar \vt^I \rho_\vm
D_\vn
\vt^J  \ .
\la{quaa} 
\ee
 Here $\r_\vm$ are projections of the 10-d Dirac
matrices, 
\be 
\r_\vm \equiv \G_{\hat m} E^{\hat m}_M \del_\vm x^M = ( \G_A
E^A_M
+\G_\apr E^\apr_M ) \del_\vm x^M \ , \ee
 and $E^{\hat m}_M$ is
the
vielbein of the 10-d target space metric.
The covariant derivative $D_\mu$ is the projection of the
10-d derivative
$D_{ M}=\del_{ M}
+ \fourth\omega^{\hat m\hat n}_{ M} \Gamma_{\hat m\hat n}
 - { 1 \ov 8 \cdot 5!}
 \G^{ \hat m_1...\hat m_5} \G_{ M}\ e^\Phi F_{ \hat m_1... \hat m_5}$
which appears in the Killing spinor equation of type
IIB supergravity, 
\be
D_\mu\t^I \equiv \left(\delta^{IJ} {\rm D}_\mu
- {\ri\ov 2 } \epsilon^{IJ} \tilde\r_\mu \right) \vt^J\ , 
\  \ \  \ \ 
\tilde\r_\mu \equiv \left(\G_A E^A_M +\ri\G_\apr
 E^\apr_M \right)
\del_\mu  x^M \ , 
\la{derr}
\ee
where $ {\rm D}_\mu = \del_\mu
+\fourth \del_\mu  x^M \omega^{\hat m\hat n}_M\Gamma_{\hat
m\hat
n}$ and      the term with   $\tilde\r_\mu$  
 originates from the coupling
 to the RR 5-form field strength (note that $F_5$ is 
 proportional to the $\epsilon$-tensors on \adss).
The presence of the  generalized ``Killing spinor"
covariant derivative explains why the action
has 32 global supersymmetries. 
Fixing the natural type IIB $\k$-symmetry gauge $\t^1=\t^2\equiv \te$ 
one finds from \rf{quaa}
\be 
S^{(2)}_{F} = 2{{i }} \int d^2 \s
 \bigg( \sqrt{g}
g^{\vm\vn}\bar \vt  \rho_\vm {\rm D}_\vn \vt 
 - { {i} \ov 2}\ep^{\mu\nu}  \bar \vt \rho_\mu {\td \rho}_\nu \te 
 \bigg)     \ .
\la{quua} 
\ee
This action  gives a well-defined  fermion kinetic term 
if one expands near generic ``long"  
string configuration. It is easy to check directly that the bosonic
\adss sigma model becomes indeed 1-loop finite when supplemented by the
fermionic action \rf{quua} \ci{DGT}
with the ``mass term" in \rf{quua} originating from the RR coupling
playing the crucial role.

The kinetic term in GS action in curved target space background
\rf{quua} involving $\te$'s 
which are 
world-volume scalars
can be transformed into the  ``2-d spinor'' form,
i.e. into the standard kinetic term for
a set of 2-d  Dirac fermions defined
on a curved 2-d space. 
In the flat space case  expanding the GS action 
near $x^0=\tau, \ x^1=\sigma$ one  concludes that 
(e.g. in the $\t^1=\t^2$ gauge)
the kinetic fermionic term takes the form 
$\bar \te \Gamma^a\del_a \te$, $a=0,1$, 
so that  choosing  the representation of
10-d  Dirac matrices where $\Gamma^{0,1}$ are 
proportional to 2-d Dirac matrices 
one  may re-interpret the action as the one for 
a collection of 2-d Dirac fermions.

In general, to achieve a similar transformation 
 one  is to apply a local
target space Lorentz rotation to GS  $\t$ (as discussed
previously mostly in flat space in  \ci{sedr}). 
 Identifying the 2-d metric 
$g_{\mu\nu }$  with  the induced metric $G_{mn} \del_\mu x^m
\del_\nu x^n$  we can write the first covariant 
derivative term in \rf{quua} as  ${ i }
\int d^2 \s
\ \sqrt {g} g^{\mu\nu }  $ $ (
 \bar\theta\r_\mu \del_\nu \theta
 - \del_\mu \bar\theta\r_\nu  \theta)
 . $ 
Introducing  the tangent $t^m_\a$ ($m=0,1,...,9$, \
$\a=0,1$) and normal $n^m_s$ ($s=1,...,8$)
vectors to the world surface which form
 orthonormal 10-d basis
($g_{\mu\nu } = e^\a_\mu  e^\b_\nu  \eta_{\a\b}$), i.e. $
t^m_\a = e^\mu _\a \del_\mu  \bar x^m , $ \ $
(t_\a, t_\b) =\eta_{\a\b}  , $  \
$ (t_\a, n_s)=0  ,$  $
(n_s, n_u)= \delta_{su} , $
where $(a,b) = G_{mn} a^m b^n$, 
 one can make a local $SO(1,9)$ rotation of this basis
which transforms the set of $\s$-dependent
10-d Dirac matrices  into the 10 constant
Dirac matrices
$
\r_\a (\s) = e^\mu_\a \r_\mu 
= S(\s) \G_\a S\inv (\s)  , $ $ 
\ 
 \r_s (\s) = n^m_s E_m^{\hat a} \Gamma_{\hat a}
 = S(\s) \G_s S\inv (\s). $  
One may further choose a representation in which
$\G_\a= \tau_\a \times I_8$, where $\tau_\a$ are 2-d Dirac
matrices.  Depending on  specific embedding
and particular curved target space metric, one may then be
able to write the action \rf{quaa} 
 as the action for 2-d Dirac
fermions coupled to curved induced 2-d metric and interacting
with some 2-d  gauge fields (coming from $S\inv dS$).

Simple examples of when this happens \ci{DGT}
will be discussed below.
We shall consider embeddings of the string world sheet into
the $AdS_3$ part of the $AdS_5$ space, so that there will be
only { one } normal direction and the extra normal bundle
2-d gauge connection will be absent.

%%%%%%%%%%%%%%%%%%%%%%%%%%%%%%%%%%%%
\section{One-loop  
partition function for  ``straight" string   }
%%%%%%%%%%%%%%%%%%%%%%%%%%%%%%%%%%%

The simplest classical solution for string in \adss
is a straight string with the world surface
spanned by the radial direction of $AdS_5$ and time.
The Euclidean solution and the corresponding induced metric
are 
\be 
x^0= \tau\,,\ \ \ 
\ 
x^4\equiv w=\sigma\,,\ \ \ \ \ \ 
\ 
ds^2  = { 1 \ov \s^2} (d\tau^2 + d\s^2)\,
. \ee 
The induced metric on the world sheet is that of $AdS_2$, \
with constant negative curvature $\R=-2$ \ (we set the  radius $R$  of
$AdS_5$  to  $1$).
This solution  corresponds 
to a single straight Wilson line at the boundary
running along the Euclidean time direction. This is a BPS
object in string theory (and  boundary gauge theory):  it
corresponds to a static fundamental string stretched between
a single D3-brane (placed at the boundary of $AdS_5$)
 and $N$ coinciding D3-branes (placed at the horizon and 
 supporting \adss 
 by their RR flux).
  Therefore,
 the partition function  should be
equal to 1. The properly defined (subtracted) classical
string action evaluated on this background  vanishes,  and the 
 corresponding 1-loop correction to the
vacuum energy defined with respect to a certain time-like
Killing vector  vanishes  too  (see below). 
Relating the vacuum energy
to the partition function using a conformal rescaling
argument one 
 concludes that $Z=1$ \ci{DGT}.\footnote{It should be mentioned that
while the properly defined vacuum energy of a
supersymmetric
field theory in $AdS$ space should vanish, this does {
not}
automatically imply (in contrast to what happens in flat
space)
that the partition function of such theory should be equal
to 1
(cf. \ci{barf,cahi}).} 
The calculation of the partition function is rather subtle,
and depends on a regularization prescription
 and proper inclusion of measure
factors.
For practical 
applications, the precise value of $Z$ (which is simply a
constant)
is not actually important, and one may normalize with
respect to
it in computing $Z$ for more general string configurations.
Any smooth Wilson loop looks in the UV region like a
straight line. This translates into the behavior of
the minimal surface near the boundary of $AdS_5$ space. In
the
general case one will have to calculate the partition
function
for a complicated two dimensional field theory. But
asymptotically the minimal surface will approach $AdS_2$,
and the
small fluctuation operators (in particular, the asymptotic
values of the masses of the fluctuation fields) will also be
the same as for a straight string. Subtleties related
to
divergences and asymptotic boundary conditions
can be automatically avoided in
more
general cases by normalizing with respect to the partition
function of the straight string.

The bosonic part of the action for small fluctuations
in conformal gauge  turns out to be  ($a,b $ are the tangent indices of
$AdS_5$ part and 
$p$ of $S^5$ part)
\be 
S_{2\rm B} =
\ha \int d^2 \s \sqrt {g}
\left(  D^\mu \z^a D_\mu  \z^a + X_{ab} \z^a \z^b
+  D^\mu  \z^p D_\mu  \z^p\right)
,\ee 
where    $
X_{ab} = {\rm diag} (1,2,2,2,1) $.
The  only non-trivial components of the 
covariant derivative   are 
$
D_0 \z^0 = \del_0 \z^0 - w^{-1}\z^4,
\
D_0 \z^4 = \del_0 \z^4 + w^{-1}\z^0.
$
Because of the direct embedding of the world sheet into the
target space the
projection
of the target space connection on the world sheet 
is
the same as the spin connection of the induced metric
and thus the
action of the conformal gauge 
ghosts is identical to the action of the longitudinal modes
$\zeta^0$, $\zeta^4$.

The quadratic part of the fermion action in the $\t^1=\t^2$ gauge 
\rf{quua} is 
\be 
L_{2\rm F} = -2i \sqrt{g} \big( \bar \theta \rho^\mu {\hat
\nabla}_\mu \theta
+ i \bar \theta \r_3 \theta \big),
\qquad
\r_3
 \equiv \ha \ep^{\a\b} \r_\a\r_\b
 = \Gamma_{04}\, , \ee
where $\r_\a=(\G_0,\G_4)$
may be identified with (2-d Dirac matrices)$\times
I_8$.
Thus the quadratic fermionic part of GS action has exactly
the
same form as the action for 2-d fermions in curved $AdS_2$ 
space with a mass term.
Assuming the standard $\int d^2 \s\sqrt{g} \bar\theta\theta$
normalization, the corresponding Dirac operator is
$ i \r^{ \mu} \hat\nabla_\mu - \r_3
= i w ( - \G_0 \del_0 + \G_4 \del_1)
 - \ha i \G_4 - \G_0 \G_4
$ 
and its square is 
$
- \hat \nabla^2 + { 1 \ov 4} \R + 1 $. 

Ignoring the ghosts and longitudinal modes\footnote{
Their contributions cancel each other modulo
 a  constant  related to difference in boundary conditions
 which is  crucial for making the full string partition function finite
 \ci{DGT}.}
  we are left with
a 2-d
field theory on $AdS_2$ containing  5 massless scalars,
3
scalars with (mass)$^2$ =2, and eight fermions with (mass)$^2$ =1.
Supersymmetric field theories on $AdS_2$ were studied, e.g., in 
 \ci{sakone,barf,iopot}.
The fields in the $\cN=1$ scalar supermultiplet in $AdS_2$
may have the following bosonic and fermionic masses
\ci{ivan,barf,sakpot}: $
m^2_B = \m^2 - \m\,,\ 
m_F = \m, $ \ ($\m$ is a free parameter).  We thus  have
5
``massless'' multiplets with $\mu=1$ ($m^2_B = 0$, $m_F =
1$)
and 3 ``massive" multiplets with $\mu=-1$ ($m^2_B = 2$, $m_F
= -1$).
It is possible to combine a $\mu=1$ and a
$\mu=-1$ multiplet into an $\cN=2$ multiplet (the
dimensional
reduction of the 4-d chiral multiplet to 2 dimensions).
Two $\mu=1$ multiplets also form an $\cN=2$ multiplet (a
 dimensional reduction of the 4-d vector multiplet).
Three chiral and one vector multiplets in $D=4$ make
one $\cN=4$ vector in four dimensions, and that suggests that
the 8 scalars and 8 fermions that we have should
form one $\N=8$ multiplet in two dimensions
(cf. \ci{sprad}).
We finally  obtain the following partition function
for the transverse and fermionic modes 
with the scalar and spinor Laplace operators defined
with respect to the Euclidean $AdS_2$ metric with radius 1 \
($\R=-2$)
\be 
Z_{B+ F}  =
{\det^{8/2}\left( - \hat \nabla^2 + \four \R + 1\right)
\over \det^{3/2}\left( - \nabla^2 + 2 \right)
\det^{5/2} \left( - \nabla^2 \right)}
\  .\ee 
 One should impose
proper boundary conditions consistent with $AdS_n$
supersymmetry \ci{avibreit}.
Those
imply that the resulting spectra of Laplace operators are
discrete
in spatial direction (and not continuous as
one would expect  in a non-compact hyperbolic space).

Instead of calculating the partition function directly, we
may start with the vacuum energy. It is given by the
determinant
of the operator scaled to remove the factor of $g^{00}$ from
in front of $\partial_t^2$ (see \ci{cahi}). 
Changing  the world sheet coordinates so that the
$AdS_2$ metric
becomes $
ds^2={\cos^{-2}\rho}\left(dt^2+d\rho^2\right), $ 
\ $  \rho\in[-{\pi\over2},{\pi\over2}]$ 
and using  the  spectra of the Hamiltonians conjugate to the  time
variable $t$  calculated in \ci{saktan}
$
\om^{(F)}_n (\m) = n + |\m| + \ha\ , \ \
$
$
\om^{(B)}_n (\m) = n + h(\m)\ ,
\ \ 
h(\m) = \ha( 1 + \sqrt{ 1 + 4m^2_B})\ , $ 
one finds, 
summing over all the modes as in \ci{sakpot},
the 1-loop vacuum energy of this effective 2-d field theory
\be 
E= \ha \sum_{n=0}^\infty\left(
3\left[\om^{(B)}_n (-1) -\om^{(F)}_n (-1)\right]
+ 5 \left[\om^{(B)}_n (1) -\om^{(F)}_n (1)\right] \right).
\ee 
The
{\it properly defined} vacuum energy should {\it vanish}
in the $AdS$ case as it does in flat space
\ci{sakone,saktan,barf,gibbo} (even though
divergences
may not cancel out, unless there is a lot of supersymmetry
\ci{ald}).
The direct computation of the sum of the mode
energies
using $\zeta$-function regularization 
gives (using the standard relations
$
\zeta(s,x) \equiv \sum_{n=0}^\infty (n+ x)^{-s},
\ 
\zeta(-1,x) = - \ha \left(x^2-x + {1 \ov 6}\right)
$) 
\be 
E
=-\fourth\left[3\times\left(2+{1\over6}\right)
+5\times {1\over6}
-8\times\left(1-{1\over12}\right)\right]
=0\,
. \ee
The ratio 3:5 of the numbers of the two multiplets
is just what is needed for the cancellation.
The fact that $E$, defined by $\zeta$-function
regularization, 
vanishes  should be a consequence of the extended $\N=8$
supersymmetry mentioned above. Indeed, while as was
found  for $AdS_4$ 
\ci{sakone,barf,gibbo}, the $\zeta$-function
regularization
may break supersymmetry and thus may lead to $E\not=0$,
this does not actually happen in the case of $N\geq 5$,
$d=4$
gauged supergravities \ci{ald}. The present $d=2$ case is thus
analogous to those $d=4$ cases with large amounts of
supersymmetry.

 In curved (e.g., static conformally flat)
space the logarithm of the 
partition function is, in general, 
 different from the vacuum energy
defined as a sum over eigen-modes because the time
derivative
part of the relevant elliptic operators is rescaled by
$g^{00}$.
The  determinants of the two operators 
which differ by such a rescaling 
are related to each other
by a  conformal anomaly type equation.
Taking this into account one finally concludes
\ci{DGT} that $Z_{tot}=1$. 
This result is a consequence of $E=0$ 
as well as  of  the cancellation of the sum of 
 conformal anomalies for ten bosons with
total mass terms 8, eight fermions with 4 times the
standard 2-d  fermion conformal
anomaly and total mass 8, and the conformal gauge  ghosts.

Similar results are  found in the case of the string world sheet
ending on a circular  Wilson loop at the boundary, 
where the induced geometry is again $AdS_2$ \ci{DGT}. 

%%%%%%%%%%%%%%%%%%%%%%%%%%%%%%%%%%%%
\section{One-loop  correction to ``bended" string \\
or rectangular Wilson loop  }
%%%%%%%%%%%%%%%%%%%%%%%%%%%%%%%%%%%
Analogous  calculation  of the one-loop correction can be 
done in the case of the ``bended" string configuration
with both ends being  at the boundary of $AdS_5$
separated by distance $L$ in a spatial direction, and 
representing  the rectangular $(T,L)$ Wilson loop 
in the boundary  gauge theory \ci{malda}.

The corresponding minimal
surface was found in \ci{malda}\ and the classical
value of the string action (equal to the minimal area 
as fermions  vanish at the classical level) 
 accounts for the
leading large  `t Hooft coupling $\l={R^4\ov \a'^2}$ behavior
 (${c_0 \sqrt \l\ov L} = { c_0 R^2 \ov \a' L} $)
of the ``quark -- anti-quark'' (W-boson) potential in the large $N$
\  
$\N=4$ SYM theory.  
On general grounds, 
in this conformal field theory  the  Wilson loop
should be $<W(C)>\ \sim\  $exp$[ - T V(L) ]$,\  $ V(L) = - {f(\l) \ov L} $, 
where  $f \to f_0 = d_1 \l + d_2 \l^2 + ...$ at weak coupling
and AdS/CFT duality  predicts \ci{malda} that 
$f \to f_\infty  = c_0  \sqrt \l + c_1 + O( {1\ov \sqrt \l}) $ 
at strong coupling.  It is natural to expect that there is 
a smooth function $f(\l)$ which interpolates between the two
expansions.
The
first  correction $c_1\ov L$ to the strong-coupling  expansion
of the potential  is determined  by the
one-loop ($\a'$) correction to the  string partition function
(see also \ci{kinar} for related discussions).

The use of the GS action of \ci{MT} as a starting point 
for  this one-loop calculation was suggested  in \ci{KT}
(were the fermionic contribution was found in the gauge $\t^1=
\G_{0123} \t^2$)
 and  it  was  completed in \ci{Forste,DGT}. 
 
 Writing the \adss metric ($R=1$)  as
 $ds^2 = y^2 dx^n dx^n + { dy^2 \ov y^2} + d \Omega_5^2$
 where $y\equiv x^4=w^{-1} $, 
 the string  configuration is described by 
 \be 
 x^0=\tau\in (0,T)\ , \ \ \ \ \ x^1 = \sigma 
  \in (-{L\ov 2},{L\ov 2})\ , \ \ \ \ \ 
 y=y(\s)\ , \ee
  where $y(\s) $ 
 extremises the action \ci{malda} 
 $S= { R^2 \ov 2 \pi \a'} T \int d\s \sqrt { y'^2 + y^4}$, i.e.
 is determined by the second-order equation
$ y y'' = 4 y'^2 + 2 y^4 $
 with the first integral being
$
y'^2 = {y^8\ov y_0^4} - y^4, $ \ 
$y_0=y_{min}= {\k_0\over L}, $ 
$  \k_0 \equiv {(2\pi)^{3/2}\ov [{\Gamma({1
\ov 4})}]^2} $.
The induced 2-d  metric takes the form
$ds^2 = y^2 d\tau^2 + y^6 d\s^2 $, \ $R^{(2)}=  - 2 ( 1 +
y^{-4})$.
 The indiced  geometry  is asymptotic to $AdS_2$:
if we change the coordinate $\s$ to $y$ the metric becomes 
\be 
ds^2 = y^2 d\tau^2 + { y^2 \ov y^4 -y_0^4} dy^2 \ ,
\ \ \ \ \ \ \ y_0 \leq y< \infty \ .\la{mee}  \ee 
The $y_0=0$ limit  corresponds to the straight string
configuration where the metric becomes that of Euclidean
$AdS_2$ space (with $0 \leq y< \infty$).

In the flat space limit ($R\to \infty$) one finds that the
quantum correction to the rectangular Wilson loop 
 vanishes because of
the cancellation of the bosonic and fermionic contributions due
to effective 2-d supersymmetry present after gauge fixing. 
The 1-loop
$c_1 \ov L$ correction to the effective potential
should  not, however, vanish \ci{KT} 
 in the present curved space case
as there is no reason to expect that the action expanded
near the solution $y=y(\s)$ should have an
effective world-sheet supersymmetry.

The bosonic fluctuations near this bended string configuration 
were analysed in \ci{Forste,DGT}. As in the straight string case 
the  one-loop
 bosonic partition function in the static gauge 
is again expressed in terms of massive and massless
scalar determinants in induced 2-d geometry.
In the static gauge \ci{Forste}
\be 
Z_{B}^{  (stat.)}
= { \det^{-2/2} \left( - \nabla^2 + 2 \right)
\det^{-1/2} \left( - \nabla^2 + \R + 4 \right)
\det^{-5/2} \left( - \nabla^2 \right)} 
\,,  \ee
while in the 
conformal gauge \ci{DGT}
\be 
Z_{B}^{  (conf.)}
= { \det^{1/2}\left( - \nabla^2_{\mu\nu } - \ha \R g_{\mu\nu} \right)
\over
\det^{1/2} \left( - D^2_{ab} + X_{ab} \right)
\det^{5/2} \left( - \nabla^2 \right) }\, ,  \ee
where $X_{ab}=2\delta_{ab} - 
g^{\mu\nu}\eta_\mu^{\,a}\eta_\nu^{\,b}\,,$ 
$ \eta_0^{\,a} = (y, 0, 0,0,0)\,,$ $
\eta_1^{\,a} = (0, y, 0,0,y\inv y')$  (we set $y_0=1$). 
The two expressions are equivalent once  supplemented with constant 
factors  related to gauge fixing. 

As discussed  above, a  systematic  way to  put the fermionic contribution into 
that of massive  2-d Dirac spinors in induced 2-d geometry
is to perform a local Lorentz rotation.
 The quadratic  part of the fermionic action depends on
 $\r_0 = y \Gamma_0\,,
\ 
\r_1 = y \Gamma_1 + y\inv y' \Gamma_4\,,
\ 
{\hat D}_\mu = \del_\mu + \ha y \Gamma_\mu \Gamma_4.$
Using  the gauge $\t^1=\t^2$  (and  Minkowski signature)
one observes \ci{DGT}  that 
 the combination of $\G_1$ and $\G_4$
which
appears in $\r_1$ can be interpreted as a (local, $\s$-dependent)
rotation of $\G_1$ in the 1-4 plane
$
S \G_1 S\inv = \cos \a\ \G_1 + \sin \a\ \G_4
= y^{-2} \G_1 + y^{-4} y' \G_4 = y^{-3} \r_1\,,
$
where
$
S= \exp \left( - {\alpha\over 2}\G_1 \G_4\right)\,,
$ $
\cos \a = y^{-2}\,, $  $
\a'\equiv {d\a\ov d \s} = 2 y\,.
$
Making the field redefinition
$
\t \to \Psi \equiv S\inv \theta\,,
$
one obtains  
\be 
L_{2\rm F} = 2i \sqrt{g} \left( \bar \Psi \tau^\mu 
 {\hat \nabla}_\mu 
\Psi
 + i \bar \Psi \tau_3 \Psi \right)
\,, \ \ \ \  
 \tau_3 \equiv {\ep^{\mu\nu} \ov 2\sqrt{g}} \tau_\mu \tau_\nu=
\G_0 \G_1\,, \ 
(\tau_3)^2 =1\, . \ee
Choosing a representation for $\G_a$ such that
$\G_{0,1}$ are $\sim$ 2-d Dirac matrices, i.e.
$\G_0 = i \s_2 \times I_8\,,
\ 
\G_1 = \s_1 \times I_8\ ,\ 
\tau_3= \G_0 \G_1 = \s_3 \times I_8,$
 we end up with 8
species of 2-d Majorana fermions living on a curved 2-d surface
with a $\s_3$ mass term. {Assuming that fermions are normalized
with $\sqrt g$}, the square of the resulting fermionic operator
is
then $
 - \hat \nabla^2 + \four \R + 1$, 
where $ \hat \nabla^2 = { 1 \ov \sqrt g} \hat \nabla^\mu 
(\sqrt g \hat \nabla_\mu )$  and 
 $\hat \nabla $ is the covariant 2-d spinor derivative
 of the induced metric.
Equivalent results for the fermionic contribution were found in  
 the ``3-brane'' gauge $\t^1= i \G_4 \t^2$ \ci{KT}  and in 
 a different $\k$-symmetry gauge in 
 \ci{Forste}.
 
Combining the contributions 
of   bosons in the static
 gauge and fermions in the $\t^1=\t^2$ gauge 
 the  total  expression for the 1-loop partition
function of a ``bended"  string in \adss\  is thus \ci{Forste,DGT}
\be 
Z
= Z_0 { \det^{8/2} ( - \hat \nabla^2 + \four \R + 1)
\over \det^{2/2} \left( - \nabla^2 + 2 \right)\ 
\det^{1/2} \left( - \nabla^2 + \R + 4 \right)\ 
\det^{5/2} \left( - \nabla^2 \right)}
\ ,  \ee
where $Z_0$ is a gauge fixing (measure)  
factor which is the same as in the flat ($R\to \infty$) limit.
In particular, this factor is crucial for cancellation \ci{DGT}
of the topological ($\log \epsilon \ \int R^{(2)}$)
divergence  present \ci{Forste}  in the ratio of the determinants.
The key point is that the 
 cancellation of this divergence in the
one-loop approximation in curved target space is
essentially the same as in the flat space, where 
to demonstrate  that logarithmic Euler number divergences cancel
(as they should to match the cancellation of conformal
anomalies in $D=10$ superstring)
one is to take into account various 
zero mode normalization factors in the string 
path integral measure \ci{alv}.\footnote{One should note also that
in performing  a local target space rotation
that transforms the quadratic GS fermion term into the 2-d
fermion kinetic term the  resulting Jacobian 
\ci{kalm} depends on the 2-d metric and its contribution
explains why the conformal anomaly of a GS fermion is 4 times
bigger than that of a 2-d fermion. This  is crucial
for understanding how conformal anomalies cancel in $D=10$
GS string.}
To avoid  questions about boundary terms (and details
of topological infinity cancellation) one  may normalize the 
partition function for each field by the partition function of
an equivalent field in the straight
string configuration, i.e. divide the partition function 
for the noncompact hyperbolic negative curvature 2-d  space \rf{mee} 
 by
the partition function for the $AdS_2$ case.\footnote{
Since the topology and the near-boundary (large $y$) behavior of
the two metrics is the same, this eliminates the problem of carefully
tracking down all boundary terms in the expressions for the
determinants and allows one  to ignore the boundary contributions
as well as the total derivative bulk terms (such as the
logarithmically divergent terms proportional to
$\int d^2 \s \sqrt g \R$).
The ratio of the determinants for the metric
\rf{mee}
 and
for its $y_0=0$ limit will be finite and well-defined.
This is the standard recipe of defining the
determinants of Laplace operators on (e.g. 2-dimensional)
non-compact spaces by using fiducial metrics of constant
negative curvature which have the same topology and asymptotic \
behavior.}
Since the indiced 2-d geometry is static, 
the calculation of the partition function  and thus of the one-loop
coefficient $c_1$ in the  potential $V(L)$ 
reduces  to determining the spectra of one-dimensional 
Schr\"odinger operators, a  tractable  problem.
A crude estimate for  the resulting coefficient $c_1$ was  discussed in
\ci{DGT}.

\section{Conclusions}

To conclude, the Green-Schwarz action for the type IIB superstrings 
in \adss  background provides a natural  and well-defined starting point
for developing $\a'$ perturbation theory  near ``long" string
configurations. The resulting partition function 
can be interpreted as that of an effective 
2-d quantum field theory 
in curved induced 2-d geometry. The one-loop string 
correction is already sensitive to details of the string coupling to RR
background and would be hard to reproduce in the standard NSR
formalism where the RR vertex operator is defined near flat space. 
One may hope that  progress towards understanding 
the spectrum of ``short"
closed strings in \adss   and thus  large $N$ 
 superconformal YM theory  for  any value of  $\l$ may be achieved 
using a  light-cone gauge approach like the one 
initiated in \ci{MET}.

\vskip 0.3 cm 

%%%%%%%%%%%%%%%%%%%%%%%%%%%%%%%%%%%%%%%%%%%%%%%%
\noindent{\bf Acknowledgements.}
%%%%%%%%%%%%%%%%%%%%%%%%%%%%%%%%%%%%%%%%%%%%%%%
I am grateful to R.R. Metsaev, N. Drukker and D.
 Gross for discussions and
collaboration.
This work was supported in part by 
the DOE grant DEFG02-91-ER-40690,  the INTAS project 991590 
and NATO grant PST.CLG 974965.

%%%%%%%%%%%%%%%%%%%%%%%%%%%%%%%%%%%%%%%%%%%%%%%%%%%%%%%%%%%%%%%%%%%%%%%%%%
%%%  The Bibliography is set using standard LaTeX macros


\begin{thebibliography}{2}
%%%%%%%%%%%%%%%%%%%%%%%%%%%%%%%%%%%%%%%
\bi{POL}
A.M.~Polyakov,
%``String theory as a universal language,''
hep-th/0006132.
%%CITATION = HEP-TH 0006132;%%
%``The wall of the cave,''
Int.\ J.\ Mod.\ Phys.\  {\bf A14}, 645 (1999)
hep-th/9809057;
%%CITATION = HEP-TH 9809057;%%
%``String theory and quark confinement,''
Nucl.\ Phys.\ Proc.\ Suppl.\  {\bf 68}, 1 (1998)
hep-th/9711002.
%%CITATION = HEP-TH 9711002;%%
\bi{mald}
J.~Maldacena,
%``The large-N limit of superconformal field theories and
%supergravity,''
Adv.\ Theor.\ Math.\ Phys.\ {\bf 2}, 231 (1998),
hep-th/9711200.   S.S.~Gubser, I.R.~Klebanov and A.M.~Polyakov,
%``Gauge theory correlators from non-critical string theory,''
Phys.\ Lett.\ {\bf B428}, 105 (1998),
hep-th/9802109.\\
E.~Witten,
%``Anti-de Sitter space and holography,''
Adv.\ Theor.\ Math.\ Phys.\ {\bf 2}, 253 (1998),
hep-th/9802150.
\bi{GS}
M.B. Green and J.H. Schwarz,
%{``Covariant description of superstrings"},
 Phys. Lett. {\bf B136}, 367 (1984).
%Nucl. Phys. {\bf B243}, 285 (1984).
\bibitem{DGT}
N.~Drukker, D.J.~Gross and A.A.~Tseytlin,
%``Green-Schwarz string in  $AdS_5 \times S^5$: 
%Semiclassical partition function,''
JHEP {\bf 0004}, 021 (2000), 
hep-th/0001204.
%%CITATION = HEP-TH 0001204;%%
\bi{HM}
M. Henneaux and L.  Mezincescu, 
%``A $\sigma$-model
%interpretation
%of Green-Schwarz covariant superstring action",
Phys. Lett. B152 (1985) 340.
\bi{MT}
R.R.~Metsaev and A.A.~Tseytlin,
%``Type IIB superstring action in $AdS_5 \times S^5$ background,''
Nucl.\ Phys.\ {\bf B533}, 109 (1998),
hep-th/9805028.
\bi{john}
J.H.~Schwarz,
%``Covariant Field Equations Of Chiral N=2 D = 10 Supergravity,''
Nucl.\ Phys.\  {\bf B226}, 269 (1983).
%%CITATION = NUPHA,B226,269;%%
\bibitem{GM1}
           M.~G\"unaydin and N.~Marcus,
%``The Unitary Supermultiplet Of N=8 Conformal Superalgebra
%Involving
%Fields Of Spin $\leq$ 2,''
Class.\ Quant.\ Grav.\  {\bf 2}, L19 (1985).
%%CITATION = CQGRD,2,L19;%%
 
\bi{KRR}
R.~Kallosh, J.~Rahmfeld and A.~Rajaraman,
%``Near horizon superspace,''
JHEP {\bf 9809}, 002 (1998)
[hep-th/9805217].
%%CITATION = HEP-TH 9805217;%%
\bi{MMT}
R.R.~Metsaev and A.A.~Tseytlin,
%``Supersymmetric D3 brane action in  $AdS_5 \times S^5$,''
Phys.\ Lett.\  {\bf B436}, 281 (1998)
[hep-th/9806095].
%%CITATION = HEP-TH 9806095;%%
\bi{pessan}
I.~Pesando,
%``A kappa gauge fixed type IIB superstring action on
%$AdS_5 \times S^5$,''
JHEP {\bf 9811}, 002 (1998),
[hep-th/9808020].
%%CITATION = HEP-TH 9808020;%%
%``All roads lead to Rome: Supersolvable and
%supercosets,''
Mod.\ Phys.\ Lett.\  {\bf A14}, 343 (1999)
[hep-th/9808146].
  \bi{kalram}
R.~Kallosh and J.~Rahmfeld,
%``The GS string action on $AdS_5 \times S^5$,''
Phys.\ Lett.\ {\bf B443}, 143 (1998),
hep-th/9808038.
%%CITATION = HEP-TH 9808038;%%
\bi{KT}
 R.~Kallosh and A.A.~Tseytlin,
%``Simplifying superstring action on $AdS_5 \times S^5$,''
JHEP {\bf 10}, 016 (1998),
hep-th/9808088.
\bi{malda}
J. Maldacena, 
%{``Wilson loops in large N field theories,''}
Phys. Rev. Lett. {\bf 80}, 4859 (1998), {hep-th/9803002};\\
S.-J. Rey and J. Yee,
%{``Macroscopic Strings as Heavy Quarks of Large N Gauge Theory and
%Anti-de Sitter Supergravity,''}
{hep-th/9803001}.
\bi{Forste}
S.~F\"orste, D.~Ghoshal and S.~Theisen,
%``Stringy corrections to the Wilson loop in N = 4 super Yang-Mills
%theory,''
JHEP {\bf 08}, 013 (1999),
hep-th/9903042; hep-th/0003068.

\bi{MTOL}
R.R. Metsaev and A.A. Tseytlin, unpublished; in proceedings of ``Strings 98''
conference,  http://www.itp.ucsb.edu/online/strings98/tseytlin.


\bi{sedr}
A.M. Polyakov, unpublished;
A.R.~Kavalov, I.K.~Kostov and A.G.~Sedrakian,
%``Dirac And Weyl Fermion Dynamics On Two-Dimensional Surface,''
Phys.\ Lett.\ {\bf B175}, 331 (1986).
%%CITATION = PHLTA,B175,331;%%
A.G.~Sedrakian and R.~Stora,
%``Dirac And Weyl Fermions Coupled To Two-Dimensional Surfaces:
%Determinants,''
Phys.\ Lett.\ {\bf 188B}, 442 (1987).
%%CITATION = PHLTA,188B,442;%%
 D.R.~Karakhanian,
%``Induced Dirac Operator And Smooth Manifold Geometry,''
preprint YERPHI-1246-32-90 (1990).
P.B.~Wiegmann,
%``Extrinsic Geometry Of Superstrings,''
Nucl.\ Phys.\ {\bf B323}, 330 (1989).
%%CITATION = NUPHA,B323,330;%%
 K.~Lechner and M.~Tonin,
%``The cancellation of worldsheet anomalies in the D=10 Green--Schwarz
%heterotic string sigma--model,''
Nucl.\ Phys.\ {\bf B475}, 535 (1996),
hep-th/9603093.
%%CITATION = NUPHA,B475,535;%%


\bi{barf}
W.A.~Bardeen and D.Z.~Freedman,
%``On The Energy Crisis In Anti-De Sitter Supersymmetry,''
Nucl.\ Phys.\ {\bf B253}, 635 (1985).
%%CITATION = NUPHA,B253,635;%%

\bi{cahi}
R.~Camporesi and A.~Higuchi,
%``Stress energy tensors in anti-de Sitter space-time,''
Phys.\ Rev.\ {\bf D45}, 3591 (1992).
%%CITATION = PHRVA,D45,3591;%%

\bi{sakone}
N.~Sakai and Y.~Tanii,
%``Supersymmetry And Vacuum Energy In Anti-De Sitter Space,''
Phys.\ Lett.\ {\bf 146B}, 38 (1984).
%%CITATION = PHLTA,146B,38;%%

\bi{iopot}
T.~Inami and H.~Ooguri,
%``Dynamical Breakdown Of Supersymmetry In Two-Dimensional Anti-De
%Sitter Space,''
Nucl.\ Phys.\ {\bf B273}, 487 (1986).
%%CITATION = NUPHA,B273,487;%%

\bi{ivan}
E.A.~Ivanov and A.S.~Sorin,
%``Wess-Zumino Model As Linear Sigma Model Of Spontaneously Broken
%Conformal And Osp(1,4) Supersymmetries,''
Sov.\ J.\ Nucl.\ Phys.\ {\bf 30}, 440 (1979);
%%CITATION = SJNCA,30,440;%%


\bi{sakpot}
N.~Sakai and Y.~Tanii,
%``Effective Potential In Two-Dimensional Anti-De Sitter Space,''
Nucl.\ Phys.\ {\bf B255}, 401 (1985).
%%CITATION = NUPHA,B255,401;%%


\bi{sprad}
J.~Michelson and M.~Spradlin,
%``Supergravity spectrum on $AdS_2 \times S^2$,''
JHEP {\bf 09}, 029 (1999),
hep-th/9906056.


\bi{avibreit}
S.J.~Avis, C.J.~Isham and D.~Storey,
%``Quantum Field Theory In Anti-De Sitter Space-Time,''
Phys.\ Rev.\ {\bf D18}, 3565 (1978).\\
%%CITATION = PHRVA,D18,3565;%%
P.~Breitenlohner and D.Z.~Freedman,
%``Stability In Gauged Extended Supergravity,''
Ann.\ Phys.\ {\bf 144}, 249 (1982).
%%CITATION = APNYA,144,249;%%



\bi{saktan}
N.~Sakai and Y.~Tanii,
%``Supersymmetry In Two-Dimensional Anti-De Sitter Space,''
Nucl.\ Phys.\ {\bf B258}, 661 (1985).
%%CITATION = NUPHA,B258,661;%%

\bi{gibbo}
C.J.~Burges, D.Z.~Freedman, S.~Davis and G.W.~Gibbons,
%``Supersymmetry In Anti-De Sitter Space,''
Ann.\ Phys.\ {\bf 167}, 285 (1986).
%%CITATION = APNYA,167,285;%%NSF-ITP-99-146
C.P.~Burgess,
%``Supersymmetry Breaking And Vacuum Energy On Anti-De Sitter Space,''
Nucl.\ Phys.\ {\bf B259}, 473 (1985).
%%CITATION = NUPHA,B259,473;%%

\bi{ald}
B.~Allen and S.~Davis,
%``Vacuum Energy In Gauged Extended Supergravity,''
Phys.\ Lett.\ {\bf 124B}, 353 (1983).

\bi{kinar}
J.~Greensite and P.~Olesen,
%``Worldsheet fluctuations and the heavy quark potential in the
%AdS/CFT  approach,''
JHEP {\bf 9904}, 001 (1999)
[hep-th/9901057];
%``Remarks on the heavy quark potential in the supergravity
%approach,''
JHEP {\bf 9808}, 009 (1998)
[hep-th/9806235];
Y.~Kinar, E.~Schreiber, J.~Sonnenschein and N.~Weiss,
%``Quantum fluctuations of Wilson loops from string models,''
hep-th/9911123.

\bi{alv}
O.~Alvarez,
%``Theory Of Strings With Boundaries: Fluctuations, Topology, And
%Quantum Geometry,''
Nucl.\ Phys.\ {\bf B216}, 125 (1983).\\
%%CITATION = NUPHA,B216,125;%%
H.~Luckock,
%``Quantum Geometry Of Strings With Boundaries,''
Annals Phys.\  {\bf 194}, 113 (1989).
%%CITATION = APNYA,194,113;%%


\bi{kalm}
R.~Kallosh and A.Y.~Morozov,
%``Green-Schwarz Action And Loop Calculations For Superstring,''
Int.\ J.\ Mod.\ Phys.\ {\bf A3}, 1943 (1988).\\
%%CITATION = IMPAE,A3,1943;%%
F.~Langouche and H.~Leutwyler,
%``Anomalies Generated By Extrinsic Curvature,''
Z.\ Phys.\ {\bf C36}, 479 (1987).
%%CITATION = ZEPYA,C36,479;%%



\bi{MET}
R.R. Metsaev and A.A. Tseytlin, hep-th/0007036.

\end{thebibliography}
\end{document}